\documentclass[article,nojss]{jss}

\usepackage{thumbpdf,lmodern}

\usepackage{framed}
\usepackage{booktabs}
\usepackage{subfigure}
\usepackage{amssymb,amsfonts}

\newcommand{\fct}[1]{\code{#1()}}

\newtheorem{definition}{\textbf{Definition}}

\newtheorem{theorem}{\textbf{Theorem}}
\newtheorem{corollary}{\textbf{Corollary}}
\newtheorem{proposition}{\textbf{Proposition}}

\author{Jian MA}
\Plainauthor{Jian MA}

\title{\pkg{copent}: Estimating Copula Entropy and Transfer Entropy in \proglang{R}}
\Plaintitle{copent: Estimating Copula Entropy and Transfer Entropy in R}
\Shorttitle{Estimating Copula Entropy and Transfer Entropy in \proglang{R}}

\Abstract{
  Statistical independence and conditional independence are two fundamental concepts in statistics and machine learning. Copula Entropy is a mathematical concept defined by Ma and Sun for multivariate statistical independence measuring and testing, and also proved to be closely related to conditional independence (or transfer entropy). As the unified framework for measuring both independence and causality, CE has been applied to solve several related statistical or machine learning problems, including association discovery, structure learning, variable selection, and causal discovery. The nonparametric methods for estimating copula entropy and transfer entropy were also proposed previously. This paper introduces \pkg{copent}, the \proglang{R} package which implements these proposed methods for estimating copula entropy and transfer entropy. The implementation detail of the package is introduced. Three examples with simulated data and real-world data on variable selection and causal discovery are also presented to demonstrate the usage of this package. The examples on variable selection and causal discovery show the strong ability of \pkg{copent} on testing (conditional) independence compared with the related packages. The \pkg{copent} package is available on the Comprehensive R Archive Network (CRAN) and also on GitHub at \url{https://github.com/majianthu/copent}.
}

\Plainkeywords{copula entropy, independence, conditional independence, variable selection, transfer entropy, nonparametric method, estimation, R}

\Address{
  Jian MA\\
  E-mail: \email{majian03@gmail.com}\\
  URL: \url{https://majianthu.github.io}\\
  ORCID: \url{https://orcid.org/0000-0001-5357-1921}
}

\begin{document}

\section{Introduction} \label{sec:intro}

Statistical independence and conditional independence are two fundamental concepts in statistics and machine learning. The research on mathematical tool for their measurement date back to the early days of the statistics discipline. The most widely used tool is correlation coefficients proposed by Pearson \citep{pearson1896mathematical}. However, it is only applicable to linear cases with Gaussian assumptions. The other popular tool for statistical independence is Mutual Information (MI) in information theory \citep{infobook}, which is defined for bivariate cases.

Copula is the theory on representation of dependence relationships \citep{nelsen2007,joe2014}. According to Sklar theorem \citep{sklar1959}, any probabilistic distribution can be represented as a copula function with marginal functions as its inputs. Based on this representation, Ma and Sun \citep{ma2008} proposed a mathematical concept for statistical independence measurement, named \textit{Copula Entropy (CE)}. They also proved the equivalence between MI and CE. CE enjoys several properties which an ideal statistical independence measure should have, such as multivariate, symmetric, non-negative (0 iff independence), invariant to monotonic transformation, and equivalent to correlation coefficient in Gaussian cases. The nonparametric method for estimating CE was also proposed in \cite{ma2008}, which is composed of two simple steps: estimating empirical copula function with rank statistic and estimating CE with the k-Nearest Neighbour (kNN) method proposed in \cite{kraskov2004}. 

Transfer Entropy (TE) (or conditional independence) \citep{schreiber2000measuring} is the fundamental concept for testing causality or conditional independence, which generalizes Granger Causality to more broader nonlinear cases. Since it is model-free, TE has great potential application in different areas. However, estimating TE is a hard problem if without assumptions. Recently, we proved that TE can be represented with CE only \citep{jian2019estimating}. According to this representation, the nonparametric method for estimating TE via CE is proposed in \cite{jian2019estimating}.

In summary, CE provides a unified theoretical framework for measuring both statistical independence and conditional independence. In this framework, statistical independence and conditional independence (causality) can be measured with only CE \citep{ma2008,jian2019estimating}. As a fundamental tool, CE has been applied to solve several basic problems, including association discovery \citep{jian2019discovering}, structure learning \citep{jian2008dependence}, variable selection \citep{jian2019variable}, and causal discovery \citep{jian2019estimating}. 

There are two similar theoretical frameworks for testing both independence and conditional independence based on kernel tricks in machine learning \citep{gretton2007nips,zhang2011uai} and distance covariance/correlation \citep{szekely2007measuring,szekely2009brownian,wang2015conditional}. Both frameworks can be considered as nonlinear generalization of traditional (partial) correlation, and both have non-parametric estimation methods. The kernel-base framework is based on the idea, called kernel mean embedding, that test correlation \citep{gretton2007nips} or partial correlation \citep{zhang2011uai} by transforming distributions into RKHS with kernel functions. The other framework is based on a concept called distance correlation defined with characteristic function \citep{szekely2007measuring,szekely2009brownian}. With this concept, Wang et al. \citep{wang2015conditional} defined a concept for conditional independence testing, called conditional distance correlation, with characteristic function for conditional function. Compared with these two frameworks, the framework based on CE is much sound theoretically due to the rigorous definition of CE and TE and much efficient computationally due to the simple estimation methods.

This paper introduces \pkg{copent} \citep{copent}, the \proglang{R} \citep{R} package which implements the nonparametric method for estimating CE and TE proposed in \cite{ma2008,jian2019estimating}, and now is available on CRAN at \url{https://CRAN.R-project.org/package=copent}. The latest release of the package is available on GitHub at \url{https://github.com/majianthu/copent}. The \pkg{copent} package in \proglang{Python} \citep{python} is also provided on the Python Package Index (PyPI) at \url{https://pypi.org/project/copent}. As the implementation of the nonparametric estimation of CE and TE, the \pkg{copent} package has great potentials in real applications of CE and TE, as demonstrated with the examples in this paper.

There are several \proglang{R} packages which implement the estimation of other popular statistical independence measures, such as \pkg{energy} for distance correlation \citep{szekely2007measuring,szekely2013energy}, \pkg{dHSIC} for multivariate Hilbert-Schmidt Independence Criterion (HSIC) \citep{gretton2007kernel,pfister2016kernel}, \pkg{HHG} for Heller-Heller-Gorfine Tests of Independence \citep{heller2013consistent,heller2016consistent}, \pkg{independence} for Hoeffding's D  \citep{hoeffding1948non} and Bergsma-Dassios T* sign covariance \citep{bergsma2014consistent}, and \pkg{Ball} for ball correlation \citep{pan2020ball}. In this paper, we will compare them with the \pkg{copent} package on the variable selection problem with real-world data in an example in Section \ref{sec:varsel}.

Several \proglang{R} packages for testing conditional independence are also available on CRAN, including \pkg{CondIndTests} on kernel-based test \citep{zhang2011uai}, and \pkg{cdcsis} on conditional distance correlation \citep{wang2015conditional}. These two packages are the implementations of the methods in the other two frameworks mentioned above. In the example on causal discovery in Section \ref{sec:cd}, we will compare them with the method for estimating TE implemented in the \pkg{copent} package.

This paper is organized as follows: the theory, estimation, and applications of CE and TE are introduced in Section \ref{s:CopEnt}, Section \ref{sec:impl} presents the implementation details of the \pkg{copent} package with an open dataset, and then three examples on simple simulation experiment, variable selection and causal discovery are presented to further demonstrate the usage of the \pkg{copent} package and to compare our package with the related packages in Section \ref{sec:examples}, and Section \ref{sec:summary} summarizes the paper.

\section{Copula Entropy}
\label{s:CopEnt}
\subsection{Theory}
\noindent
Copula theory unifies representation of multivariate dependence with copula function \citep{nelsen2007,joe2014}. According to Sklar theorem \citep{sklar1959}, multivariate density function can be represented as a product of its marginals and copula density function which represents dependence structure among random variables. This section is to define an association measure with copula. For clarity, please refer to \cite{ma2008} for notations.

With copula density, Copula Entropy is define as follows \citep{ma2008}:
\begin{definition}[Copula Entropy]
	\label{d:ce}
	Let $\mathbf{X}$ be random variables with marginals $\mathbf{u}$ and copula density $c(\mathbf{u})$. CE of $\mathbf{X}$ is defined as
	\begin{equation}
	H_c(\mathbf{X})=-\int_{\mathbf{u}}{c(\mathbf{u})\log{c(\mathbf{u})}}d\mathbf{u}.
	\label{eq:ce}
	\end{equation}
\end{definition}

In information theory, MI and entropy are two different concepts \citep{infobook}. In \cite{ma2008}, Ma and Sun proved that MI is actually a kind of entropy, negative CE, stated as follows: 
\begin{theorem}
	\label{thm1}
	MI of random variables is equivalent to negative CE:
	\begin{equation}
	I(\mathbf{X})=-H_c(\mathbf{X}).
	\end{equation}
\end{theorem}

Theorem \ref{thm1} has simple proof \citep{ma2008} and an instant corollary (Corollary \ref{c:ce}) on the relationship between information containing in joint probability density function, marginals and copula density.
\begin{corollary}
	\label{c:ce}
	\begin{equation}
	H(\mathbf{X})=\sum_{i}{H(X_i)}+H_c(\mathbf{X})
	\end{equation}
\end{corollary}
The above results cast insight into the relationship between entropy, MI, and copula through CE, and therefore build a bridge between information theory and copula theory. CE itself provides a theoretical concept of statistical independence measure.

Conditional independence is another fundamental concept in statistics with wide applications. TE is a statistical measure for causality, which is essentially conditional independence testing. It is defined by Schreiber \citep{schreiber2000measuring} as follows:
\begin{definition}[Transfer Entropy]
Let $x_t,y_t$ be two time series observations at time $t=1,\ldots,N$ of the processes $X_t,Y_t$. The transfer entropy $T_{Y \rightarrow X}$ from $Y$ to $X$ is defined as 
\begin{equation}
	T_{Y \rightarrow X} = \sum{p(x_{t+1},x_t,y_t)\log{\frac{p(x_{t+1}|x_t,y_t)}{p(x_{t+1}|x_t))}}}.
\end{equation}
\end{definition}
It can be written as conditional MI between $x_{t+1}$ and $y_t$ conditioned on $x_t$:
\begin{equation}
T_{Y \rightarrow X} = I(x_{t+1};y_t|x_t). \label{e:cmi}
\end{equation}

In \cite{jian2019estimating}, Ma proved that CE is closely related with TE, stated as the following proposition:
\begin{proposition}
	TE can be represented with only CE as follows:
\begin{equation}\label{eq:te1}
	T_{Y \rightarrow X}=-H_c(x_{t+1},x_t,y_t)+H_c(x_{t+1},x_t)+H_c(y_t,x_t).
\end{equation}
\label{tece}
\end{proposition}
This proposition can be easily proved by expanding the definition of TE \citep{jian2019estimating}. This result gives the way of measuring causality with CE. Therefore, we developed a theoretical framework for measuring both statistical independence and conditional independence with only CE.

\subsection{Estimating CE and TE}
\label{s:est}
\noindent
It is widely considered that estimating MI is notoriously difficult. Under the blessing of Theorem \ref{thm1}, Ma and Sun \citep{ma2008} proposed a non-parametric method for estimating CE (MI) from data which is composed of only two simple steps: 
\begin{enumerate}
	\item Estimating Empirical Copula Density (ECD);
	\item Estimating CE.
\end{enumerate}

For Step 1, if given data samples $\{\mathbf{x}_1,\ldots,\mathbf{x}_T\}$ i.i.d. generated from random variables $\mathbf{X}=\{x_1,\ldots,x_N\}^T$, one can easily estimate ECD as follows:
\begin{equation}
F_i(x_i)=\frac{1}{T}\sum_{t=1}^{T}{\chi(\mathbf{x}_{t}^{i}\leq x_i)},
\end{equation}
where $i=1,\ldots,N$ and $\chi$ represents for indicator function. Let $\mathbf{u}=[F_1,\ldots,F_N]$, and then one can derives a new samples set $\{\mathbf{u}_1,\ldots,\mathbf{u}_T\}$ as data from ECD $c(\mathbf{u})$. 

Once ECD is estimated, Step 2 is essentially a problem of entropy estimation which can be tackled by many existing methods. Among those methods, the kNN method \citep{kraskov2004} was suggested in \cite{ma2008}, which leads to a non-parametric way of estimating CE.

As a model-free measure of causality, TE has great potential applications in many areas. However, estimating TE is also widely considered as a hard problem. Proposition \ref{tece} presents a representation of TE with only CE. This representation suggests a method for estimating TE via CE \citep{jian2019estimating}, which composes of two steps:
\begin{enumerate}
	\item Estimating three CE terms in (\ref{eq:te1});
	\item Calculating TE with the estimated CE terms.
\end{enumerate}
Here, Step 1 is proposed to be done with the above nonparametric method for estimating CE, and hence we proposed a simple and elegant nonparametric method for estimating TE. This nonparametric method makes it possible for applying TE to real problem without any assumptions on the underlying dynamical systems.

\subsection{Applications}
CE has been applied to solve several typical statistical problems, including:
\begin{itemize}
\item{Association Measuring \citep{jian2019discovering}.} CE is used as an association measure, which enjoys many advantages over the traditional association measure, such as Pearson's correlation coefficient.
\item{Structure Learning \citep{jian2008dependence}.} Based on dependence relationship between random variables measured by CE, a graph can be derived with the maximal spanning tree algorithm.
\item{Variable Selection \citep{jian2019variable}.} For regression or classification tasks, variables can be selected based on statistical independence strength between variables and target variable measured by CE. Due to the merits of CE, such selection is both model-free and tuning-free. An example in Section \ref{sec:varsel} demonstrates the variable selection method with CE.
\item{Causal Discovery \citep{jian2019estimating}.} To discover causal relationships from observational data, transfer entropy can be estimated via CE non-parametrically to measure causality as proposed in Section \ref{s:est}. Such estimation makes no assumption on the underlying mechanism and can be applied to any cases provided time series data are available. An example in Section \ref{sec:cd} demonstrates the method.
\end{itemize}

\section{Implementation} \label{sec:impl}

The \pkg{copent} package contains five functions as listed in Table \ref{tab:functions}. The function \fct{copent} is the main function which implements the method for estimating CE and the other two functions \fct{construct\_empirical\_copula} and \fct{entknn} are called by \fct{copent} as two steps of the estimation method. The function \fct{ci} implements the method for conditional independence testing, which calls the function \fct{copent}. The function \fct{transent} implements the method for estimating TE \citep{jian2019estimating}, which calls the function \fct{ci} since estimating TE is essentially conditional independence testing.

\begin{table}[t!]
	\centering
	\caption{\label{tab:functions} The functions in the package. \code{k,dtype} represent the arguments for $k^{th}$ nearest neighbour, and distance type respectively.}
	\begin{tabular}{lp{7.4cm}}
		\hline
		Function                      & Description \\ \hline
		\code{construct\_empirical\_copula(\textbf{x})}    & constructing empirical copula function from data \code{x} based on rank statistics \\
		\code{entknn(\textbf{x},k,dtype)}            &  estimating entropy from data \code{x} with the kNN method \citep{kraskov2004} \\
		\code{copent(\textbf{x},k,dtype)}   & main function for estimating CE from data \code{x}, which is composed of two steps implemented by calling the above two functions \\ 
		\code{ci(x,y,z,k,dtype)} &testing conditional independence between \code{(x,y)} conditioned on \code{z} \\
		\code{transent(x,y,lag,k,dtype)} &estimating TE from \code{y} to \code{x} with time lag \code{lag} \\
		\hline
	\end{tabular}
\end{table}

To illustrate the implementation and usage of the \fct{copent} function, we use the ``airquality'' dataset in \proglang{R} as a working dataset, which contains daily air quality measurements in New York, May to September 1973.

\begin{CodeChunk}
\begin{CodeInput}
R> library(copent)
R> data("airquality")
R> x1 = airquality[,1:4]
\end{CodeInput}
\end{CodeChunk} 

\begin{figure}[t!]
	\centering
	\includegraphics{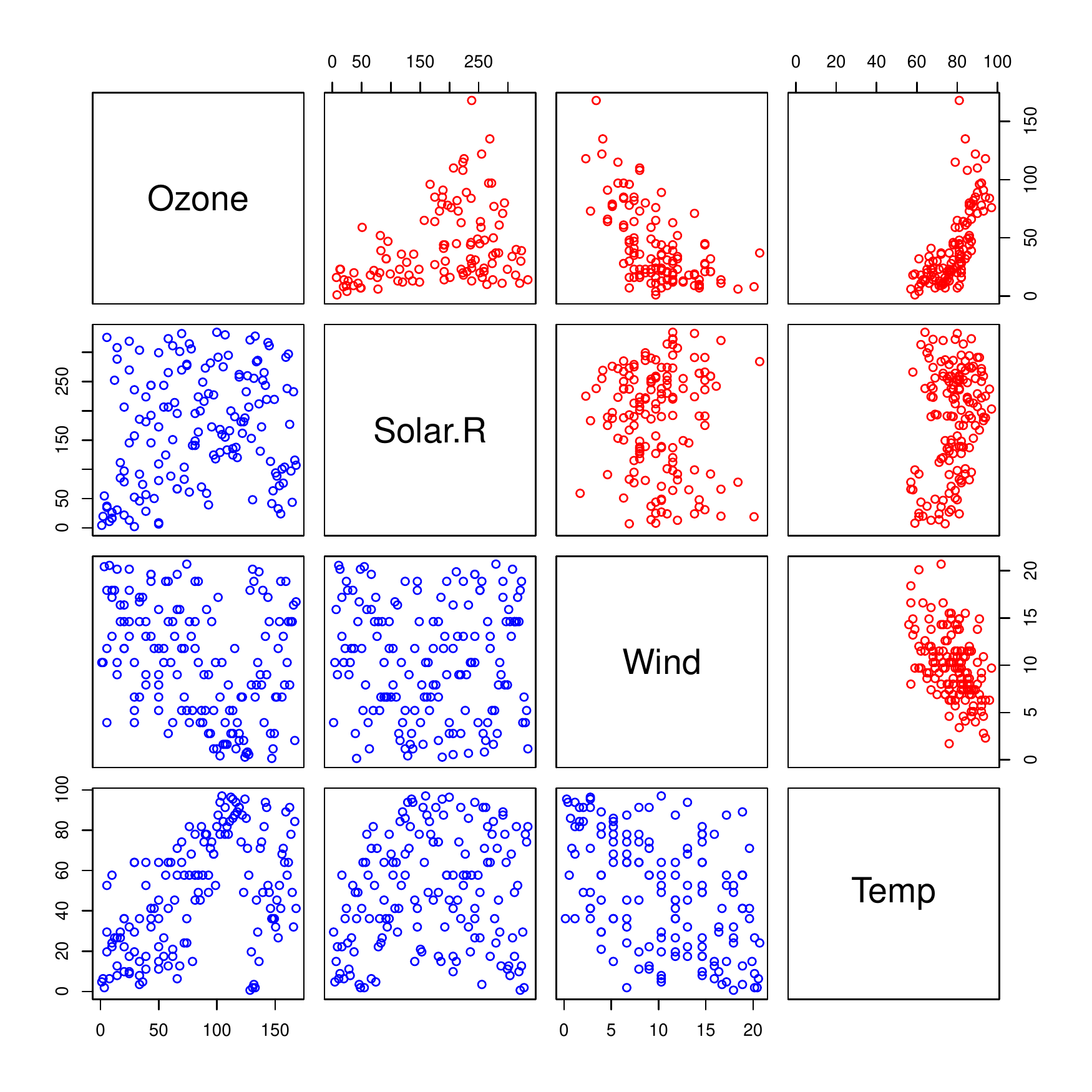}
	\caption{\label{fig:contrast} The joint distribution of the four measurements (upper panels) and the estimated empirical copula (lower panels).}
\end{figure}

The function \fct{construct\_empirical\_copula} estimates empirical copula from data with rank statistic. After the four numerical measurements are loaded, the corresponding empirical copula function can be derived by the function \fct{construct\_empirical\_copula} as follows:

\begin{CodeChunk}
	\begin{CodeInput}
R> xc1 = construct_empirical_copula(x1)
	\end{CodeInput}
\end{CodeChunk} 
The estimated empirical copula of the four measurements is illustrated in Figure \ref{fig:contrast}.

The function \fct{entknn} implements the kNN method for estimating entropy proposed in \cite{kraskov2004}. It is based on the following estimation equation:

\begin{equation}
	\hat{H}(X) = -\psi(k) + \psi(N) + \log{c_d} + \frac{d}{N}\sum_{i=1}^{N}{\log{\epsilon(i)}}.
\end{equation}
Here, $\psi()$ is the digamma function; the $c_d$ is the volume of the \textit{d}-dimensional unit ball, for which two cases are implemented: $c_d=1$ for the maximum norm and $c_d={\pi^{d/2}}/{\Gamma(1+d/2)}/{2^d}$ for Euclidean norm; and $\epsilon()$ is twice the distance from the sample to its \textit{k}-th nearest neighbor. In the package, the function \fct{entknn} has three arguments, two of which are \code{k} and \code{dtype}, \textit{k}-th neighbor and distance type (maximum norm or Euclidean norm) which are used for computing the last two terms in the above estimation equation. 

Now we can use the function \fct{entknn} to estimate the entropy of empirical copula of these four measurements:

\begin{CodeChunk}
\begin{CodeInput}
R> entknn(xc1)
\end{CodeInput}
\begin{CodeOutput}
[1] -0.03305222
\end{CodeOutput}
\end{CodeChunk}
Here we use the default value of \code{k} and \code{dtype} because of the good estimation performance of the kNN method for estimating entropy.

The main function \fct{copent} implements the method in Section \ref{s:est}. As shown above, it simply call the function \fct{construct\_empirical\_copula} to derive empirical copula function from data and then use the estimated empirical copula as input of the function \fct{entknn} to estimate CE. For user's convenience, the function \fct{copent} returns \textit{negative} value of CE. Here, the negative CE of the four measurement can be easily estimated with \fct{copent}:
\begin{CodeChunk}
\begin{CodeInput}
R> copent(x1)
\end{CodeInput}
\begin{CodeOutput}
[1] 0.03305222
\end{CodeOutput}
\end{CodeChunk}

With the main function \fct{copent}, we can easily implement the other two functions based on their theoretical relationships with CE. The function \fct{ci} for testing conditional independence \code{(x,y)} conditioned on \code{z} is implemented based on (\ref{eq:te1}) with two steps: first estimating three CEs terms of \code{(x,y,z),(y,z)} and \code{(x,z)} by calling \fct{copent} and then calculating the result from the estimated terms. Since TE is essentially conditional independence, the function \fct{transent} for estimating TE from \code{y} to \code{x} with time lag \code{lag} is then implemented as conditional independence testing with two simple steps: first preparing the data of $x_{t+lag}$, $x_t$,and $y_t$ from \code{x,y} according to \code{lag} and then call the function \fct{ci} on the prepared data according to the relationship between TE and conditional independence (\ref{e:cmi}). We will demonstrate the usage of these two functions in Section \ref{sec:cd}.

\section{Examples} \label{sec:examples}
To further demonstrate the usage and advantages of the \pkg{copent} package, three examples are presented in this section: the first one based on simulated data and the second and third one based on real-world data for variable selection and causal discovery respectively. We will compare our package with the related packages in the last two examples.

\subsection{Simulation Example}
This demonstration example is based on the simulated data. We generate the simulated data with the \pkg{mnormt} \citep{mnormt} package.

\begin{CodeChunk}
\begin{CodeInput}
R> library(copent)
R> library(mnormt)
\end{CodeInput}
\end{CodeChunk}

First, 500 data samples are generated from bivariate Gaussian distribution. Without loss of generality, the correlation coefficient $\rho$ is set as \code{0.75}.

\begin{CodeChunk}
\begin{CodeInput}
R> rho = 0.75
R> sigma = matrix(c(1,rho,rho,1),2,2)
R> x = rmnorm(500,c(0,0),sigma)
\end{CodeInput}
\end{CodeChunk}

The negative CE of bivariate Gaussian can be calculated analytically as $-\log(1-\rho^2)/2$:

\begin{CodeChunk}
\begin{CodeInput}
R> truevalue = -0.5 * log(1- rho^2)
R> truevalue
\end{CodeInput}
\begin{CodeOutput}
[1] 0.4133393
\end{CodeOutput}
\end{CodeChunk}

With the function \fct{copent}, the estimated value is:
\begin{CodeChunk}
\begin{CodeInput}
R> copent(x)
\end{CodeInput}
\begin{CodeOutput}
[1] 0.4039309
\end{CodeOutput}
\end{CodeChunk}

\subsection{Example on Variable Selection}
\label{sec:varsel}
The second example \footnote{The code for this example is available at \url{https://github.com/majianthu/aps2020}.} is about the application of the package on variable selection \citep{jian2019variable}. The data used here is the heart disease dataset in the UCI machine learning repository \citep{uci}, which contains 4 databases about heart disease diagnosis collected from four locations. The dataset includes 920 samples totally, of which only 899 samples without missing values are used in the example. Each sample is with 76 raw attributes, of which the attribute `num' is the diagnosis of patients' disease and 13 other attributes are recommended by professional clinicians as relevant \citep{nahar2013computational}. The aim of the example is to select a subset of attributes with statistical independence criteria for building the model for predicting disease status. The performance of the methods will be measured by the number of the selected attributes out of 13 recommended ones.

Besides CE, several other independence measures are also considered as contrasts in the example, including
\begin{itemize}
	\item Hilbert-Schmidt Independence Criterion (HSIC) \citep{gretton2007kernel,pfister2016kernel},
	\item Distance Correlation \citep{szekely2007measuring,szekely2013energy},
	\item Heller-Heller-Gorfine Tests of Independence \citep{heller2013consistent,heller2016consistent},
	\item Hoeffing's D Test \citep{hoeffding1948non}, 
	\item Bergsma-Dassios T* sign covariance \citep{bergsma2014consistent}, and 
	\item Ball Correlation \citep{pan2020ball}.
\end{itemize}
The \proglang{R} packages used as the implementation of the above measures are \pkg{dHSIC}, \pkg{energy}, \pkg{HHG}, \pkg{independence}, and \pkg{Ball} respectively.

The example first loads the related packages:
\begin{CodeChunk}
\begin{CodeInput}
library(copent) # Copula Entropy
library(energy) # Distance Correlation
library(dHSIC) # Hilbert-Schmidt Independence Criterion
library(HHG) # Heller-Heller-Gorfine Tests of Independence
library(independence) # Hoeffding's D test or Bergsma-Dassios T* sign covariance
library(Ball) # Ball correlation	
\end{CodeInput}
\end{CodeChunk}
And then load the data samples for the UCI repository with the following codes:
\begin{CodeChunk}
\begin{CodeInput}
scan_heart_data <-function(filename1, nl = 0){
  data1 = scan(filename1, nlines = nl, what = c(as.list(rep(0,75)),list("")))
  l = length(data1[[1]])
  data1m = matrix(unlist(data1), l, 76)
  matrix(as.numeric(data1m[,1:75]), l, 75)
}
#### load heart disease data (899 samples)
dir = "http://archive.ics.uci.edu/ml/machine-learning-databases/heart-disease/"
h1 = scan_heart_data(paste(dir,"cleveland.data",sep=""), 282*10)
h2 = scan_heart_data(paste(dir,"hungarian.data",sep=""))
h3 = scan_heart_data(paste(dir,"switzerland.data",sep=""))
h4 = scan_heart_data(paste(dir,"long-beach-va.data",sep=""))

heart1 = as.matrix( rbind(h1,h2,h3,h4) )
m = dim(heart1)[1]
n = dim(heart1)[2]	
\end{CodeInput}
\end{CodeChunk}
The above codes load the 899 samples from four datasets, of which the Cleveland dataset has only 282 samples without missing values.

With the following codes, we estimates the strength of statistical independence between 75 attributes and disease diagnosis 'num' (\#58).
\begin{CodeChunk}
\begin{CodeInput}
l = 50
ce58 = rep(0,n)
for (i in 1:n){
  for (j in 1:l){
    data2 = heart1[,c(i,58)]
    data2[,1] = data2[,1] + max(abs(data2[,1])) * 0.000001 * rnorm(m)
    data2[,2] = data2[,2] + max(abs(data2[,2])) * 0.000001 * rnorm(m)
    ce58[i] = ce58[i] + copent(data2)
  }
}
ce58 = ce58 / l
\end{CodeInput}
\end{CodeChunk}
To avoid numerical instability of the estimation algorithm, we add a slight Gaussian noise on the data. As consequents, we run the estimation for 50 times to average the uncertainty caused by the added noise. In this way, we can estimate CE from any data with both continuous and discrete values.

The code for the other measures is as following:
\begin{CodeChunk}
\begin{CodeInput}
dcor58 = rep(0,n) # Distance Correlation
dhsic58 = rep(0,n)  # Hilbert-Schmidt Independence Criterion
hhg58 = rep(0,n)  # Heller-Heller-Gorfine Tests
ind58 = rep(0,n)  # Hoeffding's D test or Bergsma-Dassios T* sign covariance
ball58 = rep(0,n) # Ball correlation
for (i in 1:n){
dcor58[i] = dcor(heart1[,i],heart1[,58])
  dhsic58[i] = dhsic(heart1[,i],heart1[,58])$dHSIC
  Dx = as.matrix(dist((heart1[,i]), diag = TRUE, upper = TRUE))
  Dy = as.matrix(dist((heart1[,58]), diag = TRUE, upper = TRUE))
  hhg58[i] = hhg.test(Dx,Dy, nr.perm = 1000)
  ind58[i] = hoeffding.D.test(heart1[,i],heart1[,58])$Dn
  #ind58[i] = hoeffding.refined.test(heart1[,i],heart1[,58])$Rn
  #ind58[i] = tau.star.test(heart1[,i],heart1[,58])$Tn
  ball58[i] = bcor(heart1[,i],heart1[,58])
}
\end{CodeInput}
\end{CodeChunk}

To compare the performance of all the measures, we check the interpretability of the selected attributes with them. The recommended variables are taken as references for the selection since they are recommended by professional researchers as clinical relevant. We choose the \#16 attribute carefully as threshold of selection for every measure's results. Such selected results of different measures are shown in Figure \ref{f:varsel}.

\begin{figure}
	\centering
	\subfigure[CE]{\includegraphics[width=0.8\textwidth]{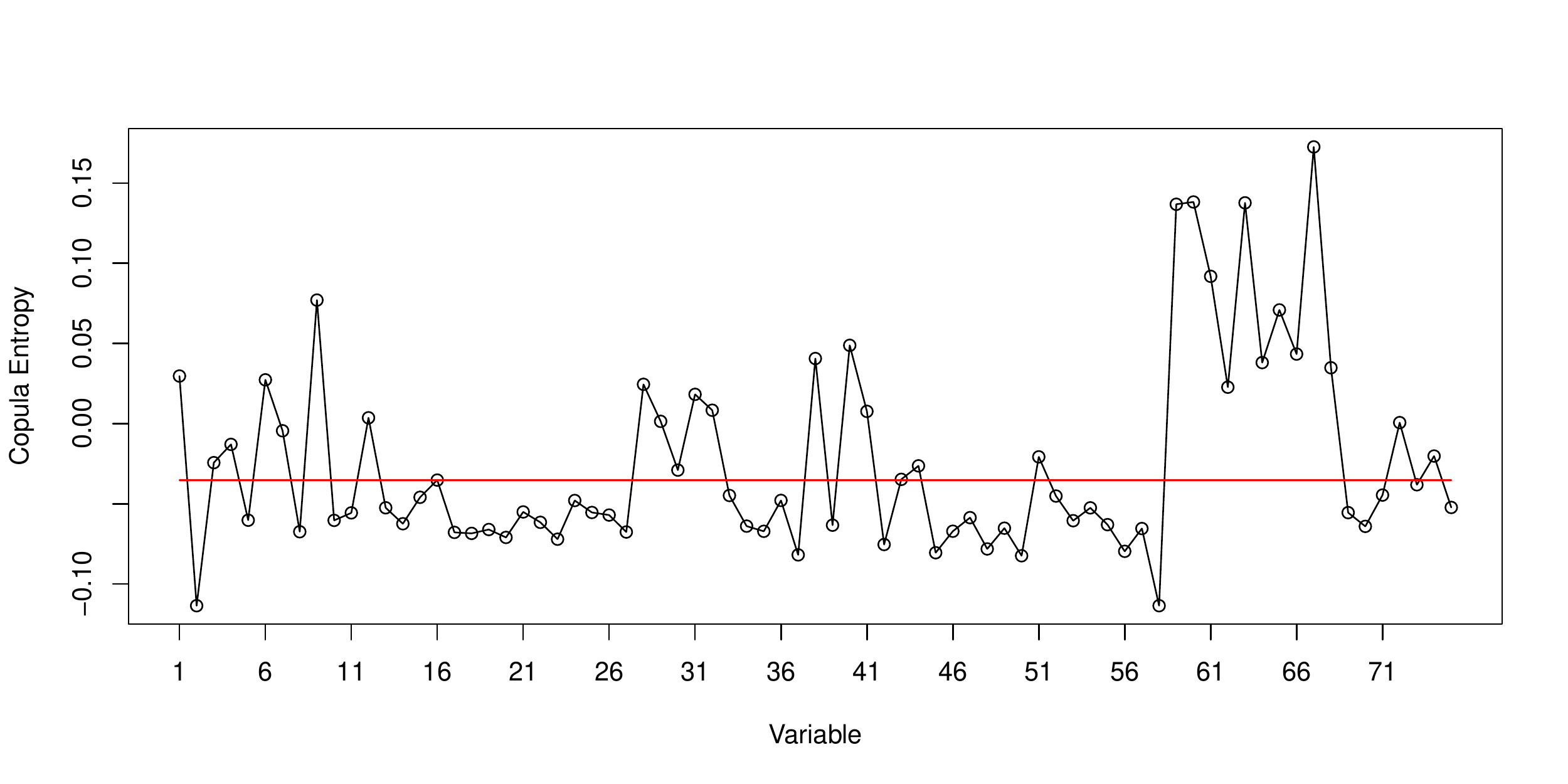}}	
	\subfigure[dCor]{\includegraphics[width=0.8\textwidth]{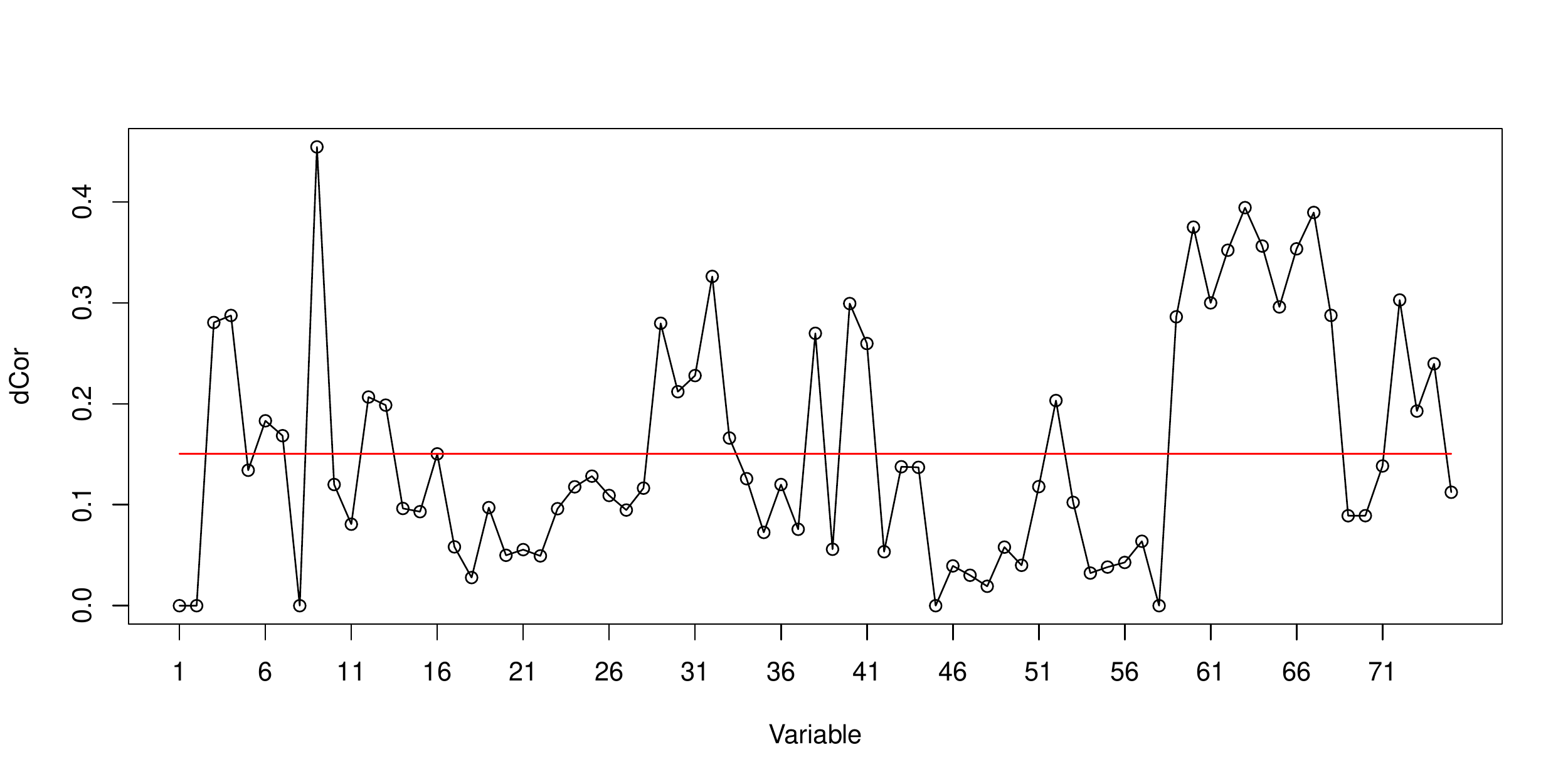}}	
	\subfigure[dHSIC]{\includegraphics[width=0.8\textwidth]{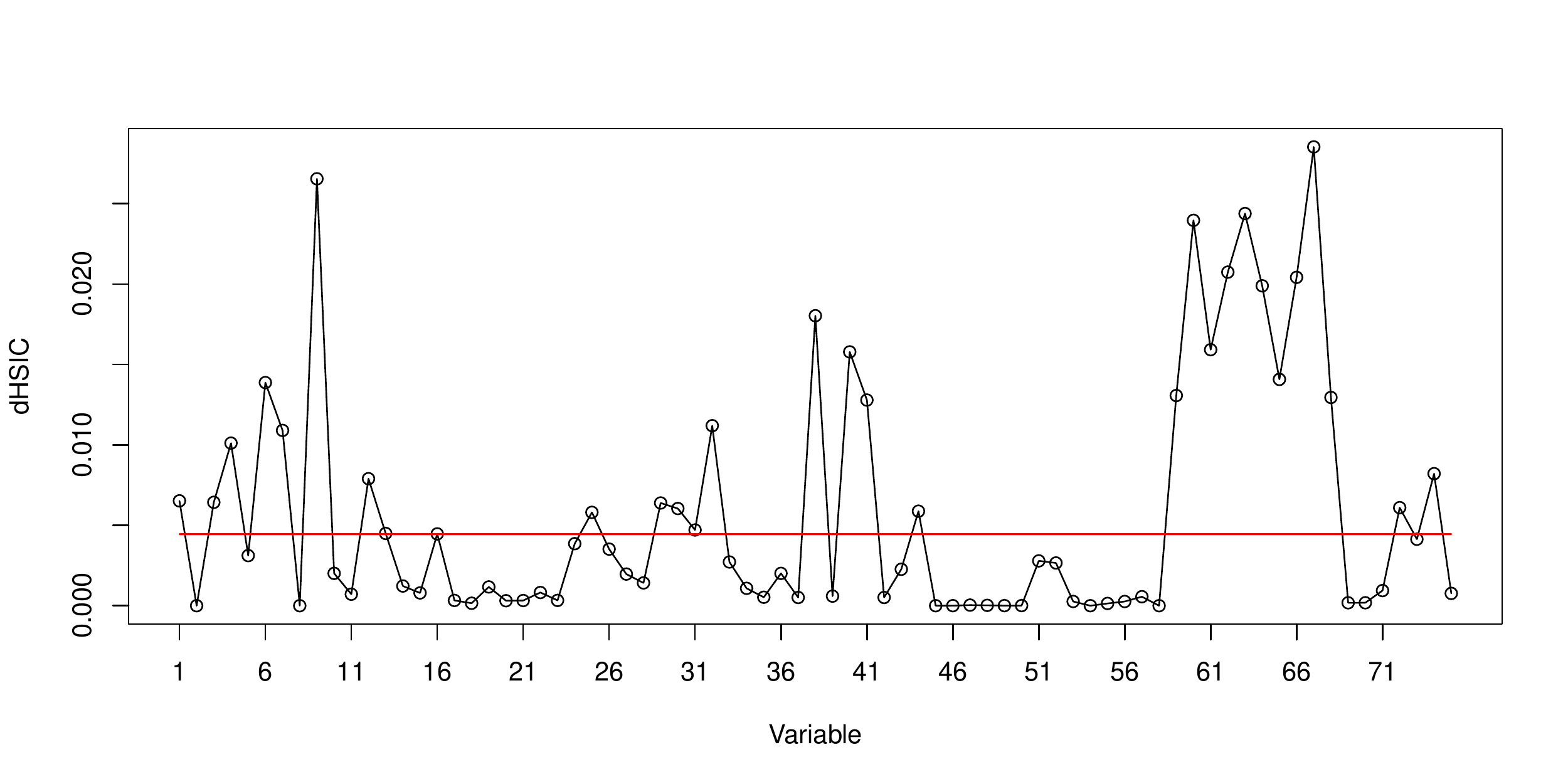}}
	\caption{Variables selected by the six dependence measures.}
	\label{f:varsel1}
\end{figure}
\addtocounter{figure}{-1}
\begin{figure}
\addtocounter{subfigure}{3}
	\centering
	\subfigure[HHG]{\includegraphics[width=0.8\textwidth]{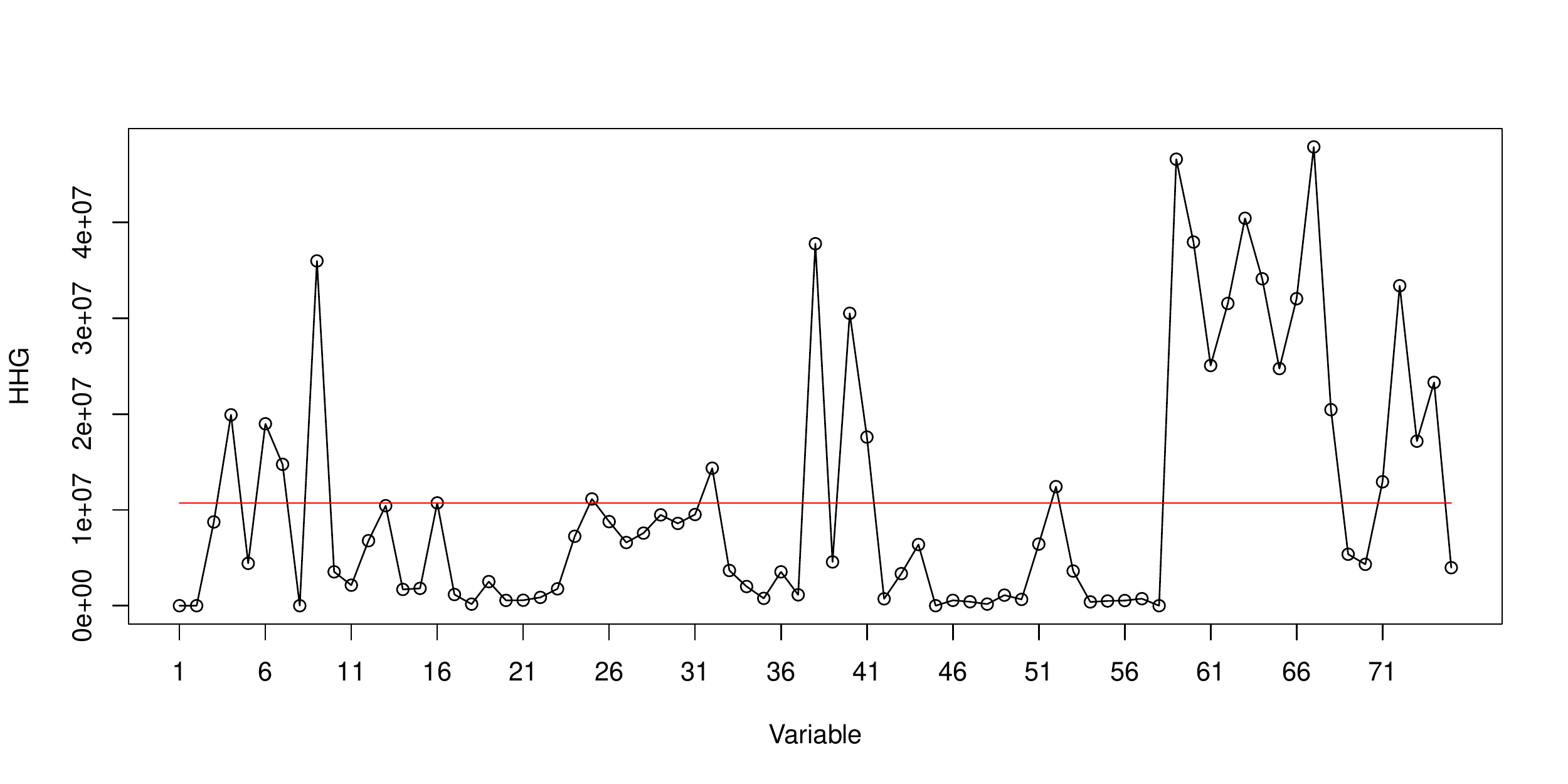}}
	\subfigure[Hoeffding's D]{\includegraphics[width=0.8\textwidth]{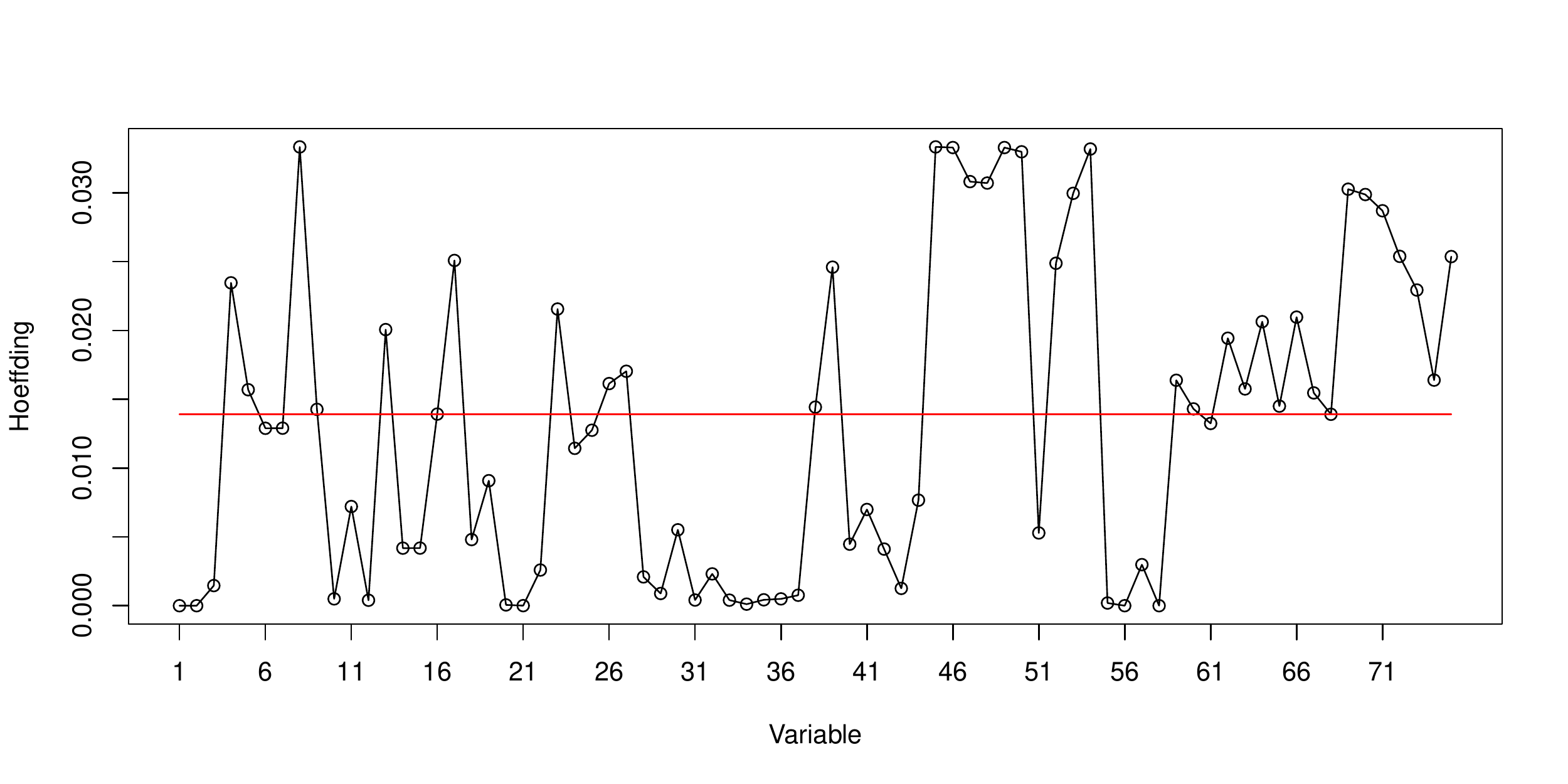}}
	\subfigure[Ball Correlation]{\includegraphics[width=0.8\textwidth]{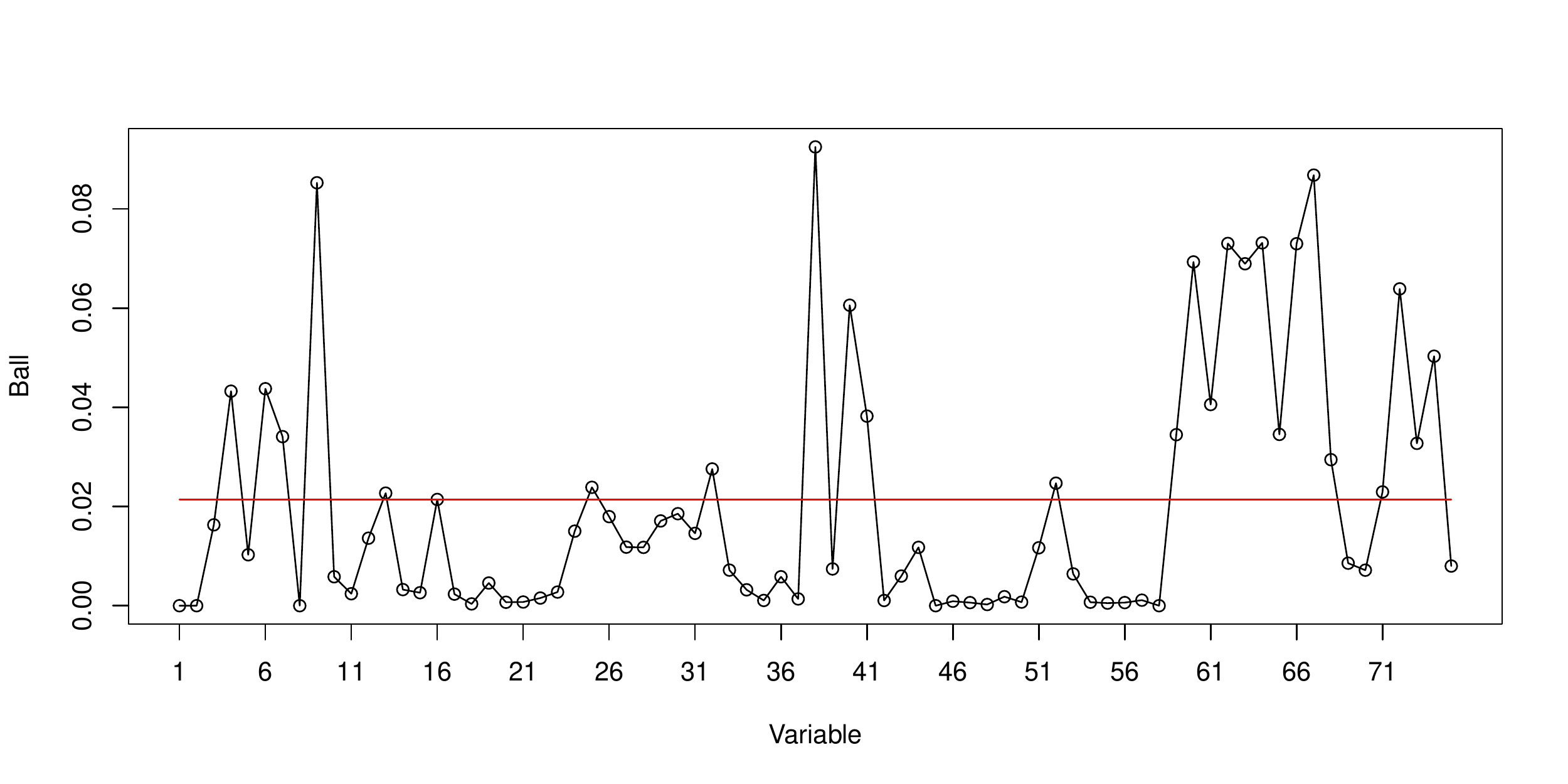}}
	\caption{Variables selected by the six dependence measures.}
	\label{f:varsel}
\end{figure}

The variables selected for different measures are summarized in Table \ref{tb:var1}. It can be learned that CE selected 11 out of 13 recommended attributes, better than all the other methods did, which means CE make much interpretable models with biomedical meaningful attributes. This result shows clearly the advantage of the \pkg{copent} package as a tool for variable selection problem.

\begin{table}
	\centering
	\caption{Selected variables by different measures.}
	\vskip2mm
	\begin{tabular}{l|c|c}
		\toprule
		\textbf{Method} & \textbf{Selected Variables' ID}& \checkmark \\
		\midrule
		Recommandation&3,4,9,10,12,16,19,32,38,40,41,44,51&13\\
		\hline
		CE	&3,4,6,7,9,12,16,28-32,38,40,41,44,51,59-68&11\\
		\hline 
		dCor &3,4,6,7,9,12,13,16,28-33,38,40,41,52,59-68&9\\
		\hline
		dHSIC &3,4,6,7,9,12,13,16,25,29-32,38,40,41,44,59-68&10\\
		\hline
		HHG &4,6,7,9,16,25,32,38,40,41,52 &7\\
		\hline 
		Hoeffding's D &4,5,8,9,13,16,17,23,26,27,38,39,45-50,52-54 &4\\
		\hline 
		Ball &4,6,7,9,13,16,25,32,38,40,41,52 &7\\
		\bottomrule
	\end{tabular}
	\label{tb:var1}
\end{table}

\subsection{Example on Causal Discovery}
\label{sec:cd}
The third example \footnote{The code for this example is available at \url{https://github.com/majianthu/transferentropy}.} is based on the Beijing PM2.5 dataset on the UCI machine learning repository \citep{uci}, which is about air pollution at Beijing. This hourly data set contains the PM2.5 data of US Embassy in Beijing. Meanwhile, meteorological data from Beijing Capital International Airport are also included. The data has been analyzed at month scale \citep{pku}. With this data, we try to discover the causal relationships between meteorological factors and PM2.5 by estimating transfer entropy via CE with the method proposed in \cite{jian2019estimating}. We also compare our method with the kernel-based methods on conditional independence testing \citep{zhang2011uai} in the package \pkg{CondIndTests}, and conditional distance correlation \citep{wang2015conditional} in the package \pkg{cdcsis}.

The example first loads the related packages:
\begin{CodeChunk}
\begin{CodeInput}
library(copent) # Copula Entropy
library(CondIndTests) # kernel based test
library(cdcsis) # conditional distance correlation
\end{CodeInput}
\end{CodeChunk}

Then the data is loaded from the UCI repository. We select only a part of data as the working set. For illustration purpose, the factors on PM2.5 and pressure are chosen in this example. Meanwhile, to avoid tackling missing values, only a continuous part of 501 hours data without missing values are used.
\begin{CodeChunk}
\begin{CodeInput}
dir = "https://archive.ics.uci.edu/ml/machine-learning-databases/00381/"
prsa2010 = read.csv(paste(dir,"PRSA_data_2010.1.1-2014.12.31.csv",sep=""))
data = prsa2010[2200:2700,c(6,9)]
\end{CodeInput}
\end{CodeChunk}

We consider causal relationship from pressure to PM2.5 with time lag from 1 hour to 24 hour. By setting time lag \code{lag} as 1 hour, we prepare the working set as follows:

\begin{CodeChunk}
\begin{CodeInput}
lag = 1
pm25a = data[1:(501-lag),1]
pm25b = data[(lag+1):501,1]
pressure = data[1:(501-lag),2]
\end{CodeInput}
\end{CodeChunk}
where \code{pm25a} and \code{pm25b} is the PM2.5 time series for `now' and `1 hour later', and \code{pressure} is the pressure time series for `now'.

%According to \cite{jian2019estimating}, the transfer entropy \textbf{TE} from $X$ to $Y$ can be represented with CE as follows:
%\begin{equation}
%	TE(X,Y) = -H_c(Y_{i+1},Y_i,X_i) + H_c(Y_{i+1},Y_i) + H_c(Y_i,X_i).
%\end{equation}

So the TE from pressure to PM2.5 with 1 hour lag can be easily estimated with the function \fct{ci} for conditional independence testing:

\begin{CodeChunk}
\begin{CodeInput}
te1[lag] = ci(pm25b,pressure,pm25a)
\end{CodeInput}
\end{CodeChunk}

Or, the estimation can also be done without preparing the working data set by simply calling the function \fct{transent} on the original data with the time lag argument \code{lag}:
\begin{CodeChunk}
\begin{CodeInput}
te1[lag] = transent(data[,0],data[,1],lag)
\end{CodeInput}
\end{CodeChunk}

The same conditional independence can also be estimated with the kernel-based and distance-based methods from the prepared data as follows:

\begin{CodeChunk}
\begin{CodeInput}
kci1[lag] = KCI(pm25b,pressure,pm25a)$testStatistic
cdc1[lag] = cdcor(pm25b,pressure,pm25a)
\end{CodeInput}
\end{CodeChunk}

\begin{figure}
\centering
\subfigure[TE via CE]{\includegraphics[width=0.8\textwidth]{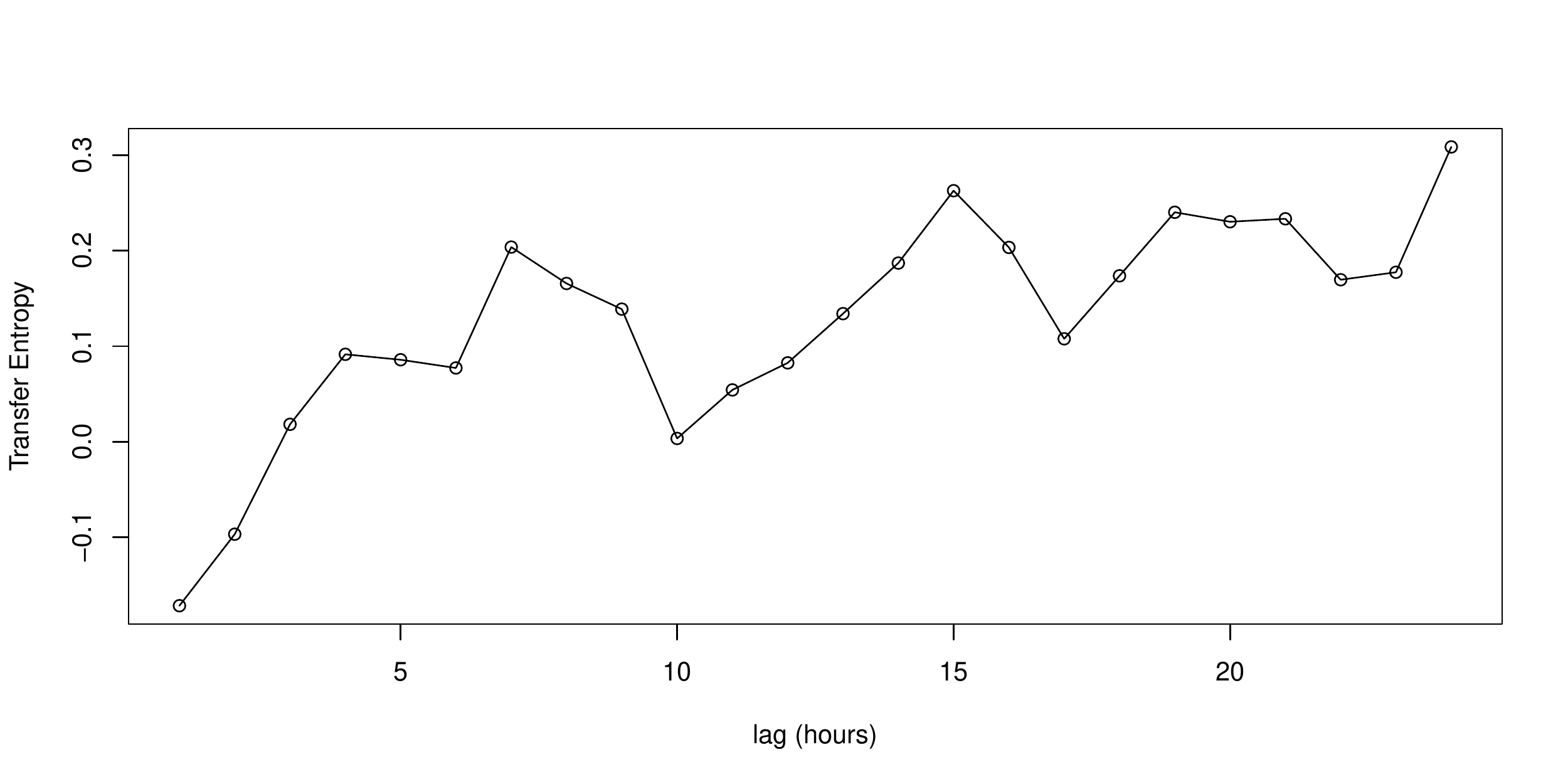}}
\subfigure[Conditional Distance Correlation]{\includegraphics[width=0.8\textwidth]{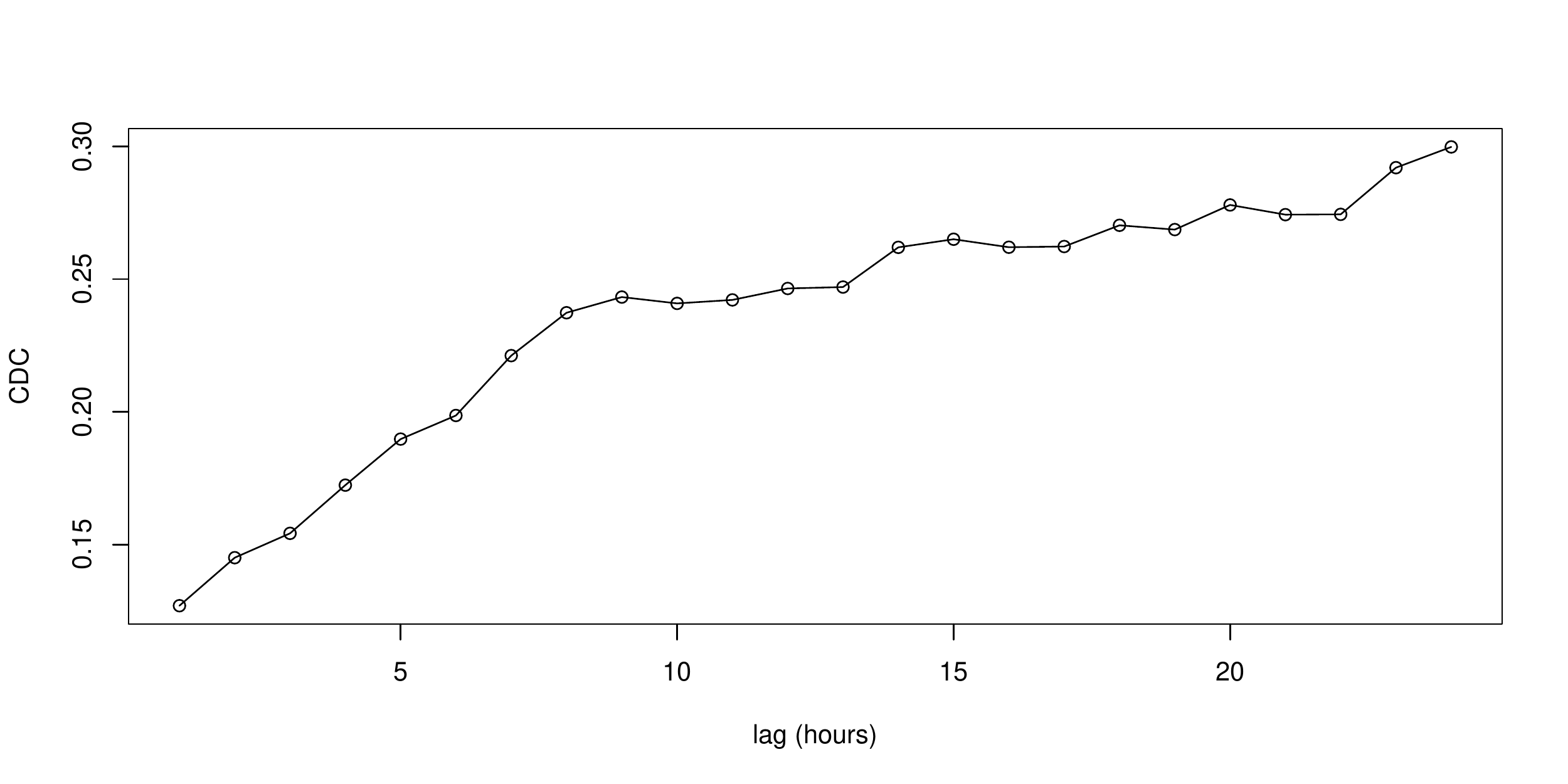}}
\subfigure[Kernel-base Conditional Independence]{\includegraphics[width=0.8\textwidth]{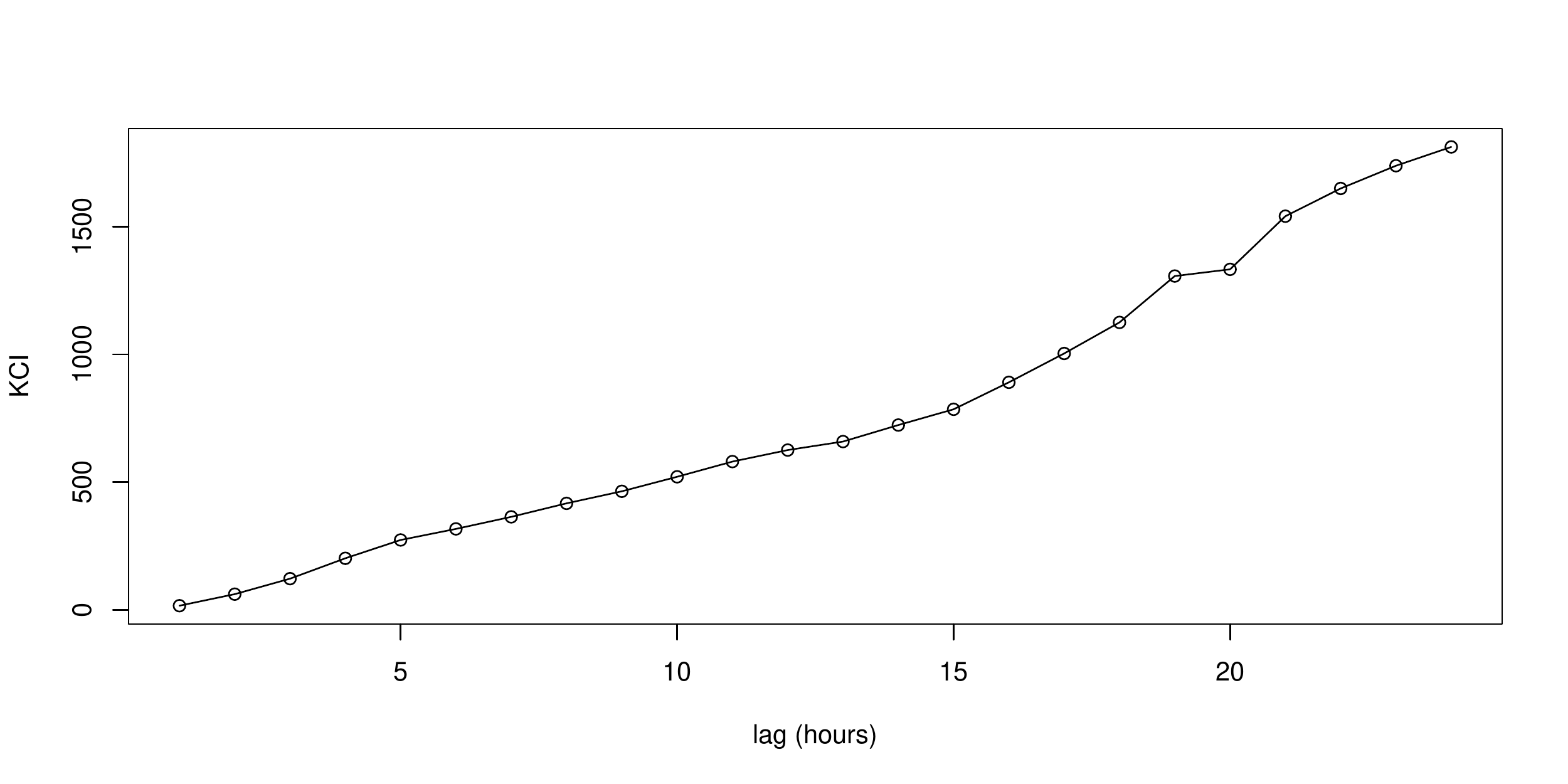}}
\caption{Three tests on causality from pressure to PM2.5 with lags from 1h to 24h.}
\label{f:causalitytests}
\end{figure}

By setting the time lag \code{lag} from \code{1} to \code{24}, we get the estimated values of three methods as illustrated in Figure \ref{f:causalitytests}. It can be learned from the Figure that the change of TE strength is well estimated which can be interpreted with meteorological meanings \citep{jian2019estimating}. Particularly, the trend of the estimated TE has mainly two phases: the sharp increasing phase and the slow increasing phase. The estimation with conditional distance correlation presents similar results as ours while the results with kernel-based method does not show such trend, which is different from ours and that of the kernel-based method.

%\begin{figure}[t!]
%	\centering
%	\includegraphics{tslag1}
%	\caption{\label{fig:tslag} The transfer entropy from pressure to PM2.5 with time lag from 1h to 24h.}
%\end{figure}

\section{Summary} \label{sec:summary}
CE is a fundamental concept for multivariate statistical independence measuring and testing and TE is a model-free concept for measuring causality. It has been proved that TE can be represented with only CE. Therefore, CE provides a unified theoretical framework for measuring both independence and conditional independence (or TE). It has been applied to solve several related statistical or machine learning problems. We have proposed the nonparametric methods for estimating CE and TE previously. In this paper, \pkg{copent}, the \proglang{R} package implementing the proposed methods, is introduced with implementation details. Three examples with simulated data and two UCI datasets on variable selection and causal discovery illustrate the usage of this package. The examples on variable selection and causal discovery show the strong ability of the \pkg{copent} package on testing (conditional) independence compared with the related packages. The \pkg{copent} package in \proglang{R} is available on the CRAN and also on GitHub at \url{https://github.com/majianthu/copent}.

\section*{Computational details}

The results in this paper were obtained using
\proglang{R}~3.6.3 with the \pkg{datasets}~3.6.3, \pkg{copent}~0.2, \pkg{energy}~1.7-7, \pkg{dHSIC}~2.1, \pkg{HHG}~2.3.2, \pkg{independence}~1.0.1, \pkg{Ball}~1.3.10, \pkg{cdcsis}~2.0.3, \pkg{CondIndTests}~0.1.5, and \pkg{mnormt}~1.5-7 packages. \proglang{R} itself
and all packages used are available from the Comprehensive
\proglang{R} Archive Network (CRAN) at
\url{https://CRAN.R-project.org/}. The UCI Heart Disease dataset and Beijing PM2.5 dataset were accessed at March 6th, 2021.

\section*{Acknowledgments}
The code of the \pkg{copent} package was first developed during the author's PhD study at Tsinghua University.

%% -- Bibliography -------------------------------------------------------------
%% - References need to be provided in a .bib BibTeX database.
%% - All references should be made with \cite, \citet, \citep, \citealp etc.
%%   (and never hard-coded). See the FAQ for details.
%% - JSS-specific markup (\proglang, \pkg, \code) should be used in the .bib.
%% - Titles in the .bib should be in title case.
%% - DOIs should be included where available.

\bibliography{refs}

\end{document}